\documentstyle[12pt,amsmath,latexsym]{article}
\def\Box{\hbox{$\sqcup$\kern-0.66em\lower0.03ex\hbox{$\sqcap$}}}

\begin{document}
\begin{titlepage}
\begin{flushright}
IFUP--TH 40/2000 \\
DFF/369/11/2000\\
\bigskip
%hep-th/xxxxxx
\end{flushright}
\vskip 1truecm
\begin{center}
\Large\bf
Hamiltonian structure and quantization of 2+1 dimensional gravity
coupled to particles 
\footnote{This work is  supported in part by M.U.R.S.T.}
\end{center}

\vskip 1truecm
\begin{center}
{Luigi Cantini$~^{a}$, Pietro Menotti$~^{b}$, Domenico Seminara}$^{c}$\\  
\end{center}
\begin{center}
\vskip 1truecm
\small\it $^{a}$ Scuola Normale Superiore, 56100 Pisa, Italy\\
e-mail: cantini@cibs.sns.it\\
{\small\it $^{b}$ Dipartimento di Fisica dell'Universit{\`a}, 56100 Pisa, 
Italy}\\
{\small\it and INFN, Sezione di Pisa}\\
e-mail: menotti@df.unipi.it\\
{\small\it $^{c}$ Dipartimento di Fisica dell'Universit{\`a}, 50125
Firenze, Italy}\\
{\small\it and INFN, Sezione di Firenze}\\
e-mail: seminara@fi.infn.it\\
\end{center}
\vskip 1truecm
\begin{center}
November 2000
\end{center}
\end{titlepage}

\begin{abstract} 
It is shown that the reduced particle dynamics of 2+1
dimensional gravity in the maximally slicing gauge has hamiltonian
form. This is proved directly for the two body problem and for the
three body problem by using the Garnier equations for isomonodromic
transformations. For a number of particles greater than three the
existence of the hamiltonian is shown to be a consequence of a
conjecture by Polyakov which connects the auxiliary parameters of the
fuchsian differential equation which solves the $SU(1,1)$ 
Riemann-Hilbert problem, to the Liouville action of the conformal
factor which describes the space-metric.

We give the exact diffeomorphism which transforms the expression of
the spinning cone geometry in the Deser, Jackiw, 't Hooft gauge to the
maximally slicing gauge. It is explicitly shown that the boundary term
in the action, written in hamiltonian form gives the hamiltonian for
the reduced particle dynamics.

The quantum mechanical translation of the two particle hamiltonian
gives rise to the logarithm of the Laplace-Beltrami operator on a
cone whose angular deficit is given by the total energy of the system
irrespective of the masses of the particles thus proving at the
quantum level a conjecture by 't Hooft on the two particle
dynamics. The quantum mechanical Green's function for the two body
problem is given.  
\end{abstract}

\section{Introduction}
\label{introd}

Gravity in 2+1 dimensions \cite{DJH} has been the object of vast
interest both at 
the classical and quantum level. Several approaches have been
pursued \cite{DJH,hooft,hooft2,wael}. In \cite{BCV,welling} the
maximally slicing gauge, or 
instantaneous York gauge, was introduced. The application of such
a gauge is restricted to universes with spacial topology of genus
$g<1$ \cite{welling,MS1}; moreover for the sphere topology it can be
applied only to the static problem \cite{MS2}. Thus the range of
applicability of such a gauge is practically 
restricted to open universes with the topology of the plane; here
however it will prove a very powerful tool.

The approach developed in \cite{BCV,welling} is first order. In
\cite{MS1,MS2} the same gauge was 
exploited in the second order ADM approach; this approach turns out to
be more straightforward than the previous one and being strictly
canonical lends itself to be translated at the quantum
level. Quantization schemes have been proposed in the absence of
particles in \cite{hooft2,hosoya2,nelsonregge,carlip} and in the
presence of particles in \cite{hooft2,matschull}.

The present paper is the continuation of two previous  
papers \cite{MS1,MS2} and goes a lot deeper into the problem.

In sect.2 we give a concise summary of the results of the previous
papers \cite{MS1,MS2}; in sect.3 we derive generalized
conservation laws starting from the time evolution of the analytic
component of the energy momentum tensor of the Liouville theory which
underlies the conformal factor describing the space metric.

In sect.4 we prove explicitly the hamiltonian nature of the reduced
particle dynamics i.e. the fact that one can give a hamiltonian
description of the time development of the system in terms of the
position and momenta of the particles. Thus this is the counterpart of
the hamiltonian description in the absence of particles for closed
universes given by Moncrief \cite{moncrief} and Hosoya and Nakao
\cite{hosoya}.

While for the two particle case the result is elementary, for three 
particles it involves the exploitation of the Garnier equations,
related to the isomonodromic transformations of a fuchsian problem. We
recall that in \cite{MS1,MS2} it was proved that such Garnier
equations are an outcome of the ADM dynamical equations of 2+1
dimensional gravity. For more than three particles the proof of the
hamiltonian nature of the reduced equations of motion and the
derivation of the hamiltonian, relies on a conjecture by Polyakov
\cite{ZT1} on the relation between the regularized Liouville action
and the accessory parameters of the $SU(1,1)$ Riemann-Hilbert
problem. Such a conjecture has been proved by Zograf and Takhtajan
\cite{ZT1} for the special cases of parabolic singularities and
elliptic singularities of finite order, but up to now a proof for
general elliptic singularities is absent.

In sect. 5 we give the exact diffeomorphism which relates the conical
metric of Deser, Jackiw and 't Hooft (DJH) in the presence of angular
momentum to its description in the 
maximally slicing gauge; as a by-product it gives the exact relation
between the asymptotic metrics in the DJH and in the maximally
slicing gauge. These results will be useful in the following to
understand the boundary terms in the action. We write also the exact
expression of the Killing vectors in the maximally slicing gauge.

In sect.6 we connect the results of sect.4 with the boundary terms of
the gravitational action; 2+1 dimensional gravity coupled to particles
is an example in which one can compute the hamiltonian explicitly as a
boundary term. The dynamics is described completely by such boundary
terms of the action.

Finally in sect.7 we treat the quantization of the two particle
problem starting form the classical two particle hamiltonian. The
quantum hamiltonian turns out to be the logarithm of the
Laplace-Beltrami operator on a cone whose aperture is given by the
total energy of 
the system and is independent of the masses of the two particles. This
provides a complete proof of the conjecture of 't Hooft \cite{hooft3}
about the two particle dynamics, {\it i.e.} the equivalence of the 
relative motion of two particles with that of a test particle on 
a cone of aperture equal to the total energy. Obviously, the 
ordering problem is
always present but the Laplace-Beltrami operator appears to be the
most natural choice. A very similar structure was found and
thoroughly examined by Deser and Jackiw \cite{deserjackiw}, when
treating the quantum 
problem of a test particle moving on a cone; the main difference is
that in the present treatment its logarithm rather than the Laplace-
Beltrami operator appears. 

Given the hamiltonian one can easily compute the Green function; it
can be written in terms of hypergeometric functions. 
 
The quantum mechanical problem with more than two particles requires a
more explicit knowledge of the hamiltonian which is related to the
auxiliary parameters $\beta_B$. The existence of those parameters is
assured by the solvability of the Riemann-Hilbert problem and one can
try to produce a perturbative expansion of them at least in some
limit situations. Here however the ordering problem is likely to be
more acute.

\section{Hamiltonian approach}

To make the paper relatively self-contained we shall summarize in this
section some results of the papers \cite{MS1,MS2}.
With the usual ADM notation for the metric \cite{ADM}
\begin{equation}
ds^2 = -N^2 dt^2+ g_{ij}(dx^i+ N^i dt)(dx^j+ N^j dt)
\end{equation}
the gravitational action expressed in terms of the
canonical variables is \cite{hawkinghunter,wald,MTW}
\begin{multline} 
S_{Grav} = \int dt \int_{\Sigma_t} d^Dx
\left[ \pi^{ij}\dot g_{ij} - N^i H_i - NH\right]+ \\
+2\int dt \int_{B_t} d^{(D-1)}x \,\sqrt{\sigma_{Bt}} N 
\left( K_{B_t}+\frac{\eta}{\cosh\eta}{\cal D}_\alpha v^\alpha\right)
-2\int dt \int_{B_t} d^{(D-1)}x\, r_\alpha
\pi^{\alpha\beta}_{(\sigma_{Bt})} N_\beta. 
\end{multline} 
where $\sinh\eta=n_\mu u^\mu $ with $n^\mu $ the future
pointing unit  
normal to the time slices $\Sigma_t$ and $u^\mu $ the outward pointing
unit normal to space-like boundary $B$; 
$B_t= \Sigma_t\cap B$, $\sqrt{\sigma_{Bt}}$ stands for the
volume form induced by the 
space metric on $B_t$, $K_{B_t}$ is the extrinsic
curvature of $B_t$ as a surface embedded in 
$\Sigma_t$, $v^\alpha\equiv \displaystyle{\frac{1}{\cosh\eta}\left 
(n^\alpha-\sinh\eta~ u^\alpha\right)}$ and $r_\alpha$ is the versor
normal to $B_t$ in $\Sigma_t$.    
The subscript $\sigma_{Bt}$ in $\pi^{\alpha\beta}_{(\sigma_{Bt})}$ is
a reminder  
that it has to be considered a tensor density with respect to
the measure $\sqrt{\sigma_{Bt}}$. 
The explicit form of $H$ and $H_i$ can be found in \cite{MS1}

\noindent
The matter action can be rewritten as 
\begin{equation} 
S_m=\int\!d t \sum_n\Big(P_{ni}\, \dot
q_n^i+N^i(q_n) P_{ni} - N(q_n) \sqrt{P_{ni} P_{nj} g^{ij}(q_n)+
m_n^2}\Big). 
\end{equation}
In the $K=0$ gauge and using the complex coordinates $z = x + i y$ the
diffeomorphism constraint is simply solved by 
\begin{equation}\label{pibarzz}
\pi^{\bar z}_{~z} = -\frac{1}{2\pi}\sum_n\frac{P_n}{z-z_n}
\end{equation}
subject to the restriction $\sum_n P_n =0$ \cite{MS2}.
Always for $K=0$ and using the conformal gauge for the space metric
i.e.
\begin{equation}\label{ADMmetric}
ds^2= -N^2 dt^2+ e^{2\sigma}(dz+ N^z dt)(d\bar z+ N^{\bar z} dt)
\end{equation}
the hamiltonian constraint takes the form of the
following inhomogeneous Liouville equation  
\begin{equation}
\label{eqsigmatilde2}
2\Delta\tilde\sigma=-e^{-2\tilde\sigma}-4\pi \sum_n \delta^2(z- z_n)(
\mu _n -1)-4\pi\sum_A \delta^2(z- z_A).    
\end{equation}
In eq.(\ref{eqsigmatilde2}) $\tilde\sigma$ is defined by
\begin{equation}
e^{2\sigma} = 2 \pi^{\bar z}_{~z} \pi^{z}_{~\bar z} e^{2\tilde\sigma},
\end{equation}
$\mu_n$ are the particle masses divided by $4\pi$, $z_n$ the
particle positions and $z_A$ the positions of the $({\cal N}-2)$ apparent
singularities i.e. of the zeros of eq.(\ref{pibarzz}). 
The Lagrange multipliers $N$ and $N^z$ where expressed in terms of
$\tilde\sigma$ through 
\begin{equation}\label{N}
N = \frac{\partial(-2\tilde\sigma)}{\partial M}
\end{equation}
and
\begin{equation}\label{Nz}
N^z =-\frac{2}{\pi^{\bar z}_{\ z}(z)} \partial_z N +g(z),
\end{equation}
with  
\begin{equation}\label{generalg}
g(z) = \sum_B \frac{\partial\beta_B}{\partial M} \frac{1}{z-z_B} 
\frac{{\cal P}(z_B)}{\prod _{C\neq B} (z_B-z_C)} + p_1(z) 
\end{equation}
and  ${\cal P}$ is defined by
\begin{equation} 
-\frac{\pi^{\bar z}_{~z}(z)}{2}= \frac{1}{4\pi}
\sum_n\frac{P_{nz}}{z-z_n}\equiv\frac{\prod_B(z-z_B)}{{\cal P}(z)}. 
\end{equation}
$p_1(z)= c_0(t) + c_1(t) z$ is a first order polynomial. The role
of the first term 
in $g(z)$ is to cancel the poles arising in the first term of
eq.(\ref{Nz}) due to the zeros of $\pi^{\bar z}_{~z}$ and $\beta_B$
are the accessory parameters of the 
fuchsian differential equation \cite{kra} which underlies the solution
of the Liouville equation (\ref{eqsigmatilde2}).   
The equations for the particle motion are given by \cite{MS1}
\begin{equation}\label{dotz}
\dot z_n = - N^z(z_n) = -g(z_n) = -\sum_B
\frac{\partial\beta_B}{\partial M}\frac{1}{z_n-z_B} 
\frac{{\cal P}(z_B)}{\prod _{C\neq
B} (z_B-z_C)} - p_1(z_n)
\end{equation}
$$
\dot P_{n z} = 4\pi \frac{\partial \beta_n}{\partial M}+P_{nz} ~
g'(z_n)=
$$
\begin{equation}\label{dotP}
=4\pi \frac{\partial \beta_n}{\partial M} - P_{nz}\sum_B
\frac{\partial \beta_B}{\partial M} \frac{{\cal P}(z_B)}{(z_n-z_B)^2
\prod_{C\neq B}(z_B-z_C)}
+ P_{nz} ~p'_1(z_n).
\end{equation}
If we want a reference frame which does not rotate at infinity the
linear term in $p_1(z)$ must be chosen so as to cancel in $N^z$ the
term increasing linearly at infinity; such a choice is unique and
given by $-z/(\sum_n P_n z_n)$.

In the simple two particle case one obtains the equations of motion in
the relative coordinates $z'_2 = z_2 -z_1$, $P'=P_2 = -P_1$ 
\begin{equation}\label{system}
\dot z'_2 =\frac{1}{P'_z};~~~~\dot P'_z = - \frac\mu {z'_2}.
\end{equation}
It is interesting that in order to reach eq.(\ref{system}) there is no
need to solve the Liouville equation; the local properties of the
fuchsian differential equation underlying eq.(\ref{eqsigmatilde2}) are
sufficient. The solution of eq.(\ref{system})
\begin{equation}\label{trajectory}
z'_2 = {\rm const}~ [(1-\mu )(t-t_0) - iL/2]^{\frac{1}{1-\mu }}
\end{equation}
agrees with the solution found in \cite{BCV}. A still simpler
derivation of eq.(\ref{trajectory}) as a ratio of two conservation
laws, will be given in the next section.

\section{Virasoro generators and conservation laws}

In ref.\cite{MS1} the following generalized conservation law (and its
complex conjugate) for the ${\cal N}$ particle problem was obtained
\begin{equation}\label{dilconserv}
\sum_n P_n z_n = (1-\mu)(t-t_0) - iL 
\end{equation}
by using the particle equations of motion eq.(\ref{dotz}, \ref{dotP}),
where $4\pi \mu$ is the total energy of the system and $L$ the
angular momentum.  

In the two particle case eq.(\ref{dilconserv}) is simply $P' z'_2
= (1-\mu)(t-t_0) - iL$. As can be easily checked the hamiltonian for
eqs.(\ref{system}) and 
their complex conjugates is given by the sum of two conserved
hamiltonians i.e. $H = h + \bar h$ with $h= \ln(P' {z'}_2^\mu)$. Taking
the ratio of $P' {z'}_2^\mu = \exp( h ) = {\rm const.}$ with the
previous equation we obtain the solution eq.(\ref{trajectory}) without
the need to solve the system (\ref{system}). 

In this section we want to give a treatment of these and analogous
conservation laws from a more general viewpoint.

In ref.\cite{MS1} the following equation was derived from the ADM
formalism, with regard to the time evolution of the function $Q(z)$
appearing in the fuchsian differential equation
\begin{equation}\label{Qequation}
\dot Q(z) = \frac{1}{2}g'''(z) + 2 g'(z) Q(z) +g(z) Q'(z).
\end{equation}
$Q(z)$ can be understood as the analytic component of the energy
momentum tensor of the Liouville theory governing the conformal 
factor $\tilde \sigma$ and the above equation
represents the change of this anomalous energy momentum tensor 
under a conformal transformation generated by $g(z)$. 
It was also shown in ref.\cite{MS1} that eq.(\ref{Qequation}) contains
all the dynamics of the system, i.e. the motion of the particle
singularities and auxiliary singularities and the change in time of
the residues at such singularities; it provides also an interpretation
of 2+1 dimensional gravity \'a la Einstein-Infeld-Hoffmann \cite{EIH}.
Following a well trodden path, we want now to convert
eq.(\ref{Qequation}) into equations for the Laurent series
coefficients of $Q(z)$. With  
\begin{equation}
\frac{1}{2\pi i}\oint z^{n+1} Q(z) dz = L_{n}
\end{equation}
we obtain
\begin{equation}
L_{-1} = \frac{1}{2}(\sum_n\beta_n +\sum_B \beta_B)
\end{equation}
\begin{equation}
L_{0} = \frac{1}{2}(\sum_n\beta_n z_n +\sum_B \beta_B z_B)+
        \frac{1}{4}\left[\sum_n (1-\mu_n^2) -3 ({\cal N}-2)\right] 
\end{equation}
\begin{equation}
L_{1} = \frac{1}{2}(\sum_n\beta_n z^2_n +\sum_B \beta_B z^2_B)+
        \frac{1}{2}(\sum_n (1-\mu_n^2) z_n -3 \sum_B z_B) 
\end{equation}
and the following equation of motion
\begin{equation}\label{L-1}
\dot L_{-1} = \frac{c_1}{2}(\sum_n\beta_n +\sum_B \beta_B) 
\end{equation}
\begin{equation}\label{L0}
\dot L_{0} = -\frac{c_0}{2}(\sum_n\beta_n +\sum_B \beta_B) 
\end{equation}
\begin{equation}\label{L1}
\dot L_{1} = -c_0 L_0 -c_1 L_1.
\end{equation}
In this paper we shall restrict ourselves to $L_{-1}, L_{0}, L_{1}$. We
recall that $c_0$ and $c_1$ are function of time which specify the 
translations and roto-dilatations of the reference frame.
Eq.(\ref{L-1}) is simply a consistency requirement on the first Fuchs
relation $\sum_n \beta_n + \sum_B \beta_B =0$.
Eq.(\ref{L0}) tells us through the first Fuchs relation that $L_0$ is
constant. The value of the constant is actually provided by the second
Fuchs relation
\begin{equation}
4 L_0 = 2 \sum_n \beta_n z_n + 2 \sum_B \beta_B z_B+
\sum_n (1-\mu_n^2) -3 ({\cal N}-2) = 1 - \mu_\infty^2 = 1 - (1-\mu)^2.
\end{equation}
This implicitly shows that the total mass $\mu$ is constant in time and
more importantly, by taking the derivative with respect to $\mu$, we have
\begin{equation}
1-\mu = \sum_n\frac{\partial \beta_n}{\partial \mu} z_n
+\sum_B\frac{\partial \beta_B}{\partial \mu} z_B  
\end{equation}
which combined with the equations of motion provides the generalized
conservation law, obviously related to the dilatations
\begin{equation}\label{dilatation}
\frac{d}{dt}(\sum_nP_n z_n) = 1-\mu.
\end{equation}
We notice that due to $\sum_n P_n=0$, $\sum_n P_n z_n$ is invariant
under translations, in addition to rotations and dilatations. 
Eq.(\ref{L1}) for the time evolution of $L_1$, keeping in mind that
$L_0$ is a constant is easily solved in the form 
\begin{equation}
L_1(t) = - L_0\int_0^t c_0(t') dt'  \exp(\int_0^{t'} c_1(t'') dt'')
\exp(-\int_0^{t} c_1(t') dt') + \kappa \exp(-\int_0^{t} c_1(t') dt'). 
\end{equation}
We want now to translate such information on the physical variables
$z_n$, $P_n$.
By using the equation of motion we obtain
\begin{equation}\label{L11}
\frac{d}{dt}(\sum_nP_n z^2_n) = -2 c_0 \sum_n P_n z_n - c_1 \sum_nP_n
z^2_n  + \sum_n \frac{\partial \beta_n}{\partial \mu}z_n^2  + \sum_B
\frac{\partial \beta_B}{\partial \mu} z_B^2.
\end{equation}
The time development of
\begin{equation}
 \sum_n \frac{\partial \beta_n}{\partial \mu}z_n^2  + \sum_B
\frac{\partial \beta_B}{\partial \mu} z_B^2  
\end{equation}
in eq.(\ref{L11}) depends on the functions $c_0(t)$, $c_1(t)$.
Let us consider first the (rotating) frame defined by
$c_0=c_1=0$. Then from eq.(\ref{L1}) we have $\dot L_1 =0$. Taking  the
derivative of that equation with respect to $\mu$ we have
\begin{equation}
\frac{d}{dt}\left(\sum_n \frac{\partial \beta_n}{\partial \mu}z_n^2  
+ \sum_B \frac{\partial \beta_B}{\partial \mu} z_B^2\right) =0
\end{equation}
and thus in such a reference frame
\begin{equation}
\sum_n P_n z^2_n = At +B.
\end{equation}
Let us now consider the non rotating frame given by $c_1(t) =
-[(1-\mu)t-ib]^{-1}$ and let us consider the case $c_0(t)=0$. We
have
\begin{equation}
\frac{\partial L_1}{\partial \mu} =  \frac{1}{2}\left(\sum_n \frac{\partial
\beta_n}{\partial \mu} z_n^2  + \sum_B \frac{\partial \beta_B}{\partial 
\mu} z_B^2\right) = k_1 [(1-\mu)t-ib]^{\frac{1}{1-\mu}}. 
\end{equation}
This result can be substituted in eq.(\ref{L11}) which solved gives
\begin{equation}\label{trans}
\sum_n P_n z^2_n = (k_2 t+k_3)[(1-\mu)t-ib]^{\frac{1}{1-\mu}}
\end{equation} 
where $k_2, k_3$ are constants.
We notice that the non rotating frame with $c_0(t)=0$ corresponds to
an asymptotic behavior for $N^z$ given by
\begin{equation}
N^z = z (z\bar z)^{\mu-1} \ln(z\bar z)\left (1
+O\left(\frac{1}{|z|}\right)\right). 
\end{equation} 
Since the l.h.s. of eq.(\ref{trans}) is not translation invariant,
it provides, once the relative motion of the particles has been
solved, information on the overall motion of the system e.g. on
$z_1$.

\section{Hamiltonian nature of the reduced dynamics}

In \cite{MS1} starting form the ADM action in the presence of particles 
we have reached the particle equations of motion in the maximally
slicing gauge $K=0$ eqs.(\ref{dotz},\ref{dotP}).
As we followed a canonical procedure, we expect equations
(\ref{dotz},\ref{dotP}) to be canonical 
i.e. derivable from a hamiltonian. The present section is devoted to
the direct proof that such equations are indeed canonical i.e. are
generated by a hamiltonian and to the construction of such hamiltonian.

\noindent
To start, by means of the transformation of generator
\begin{equation}
G(z,\tilde P) =\sum_n(a_1(t)z_n+a_0(t)) \tilde P_n
\end{equation}
with
\begin{equation}
c_1(t) = - \frac{\dot a_1(t)}{a_1(t)}; ~~~~ c_0(t) = - \dot a_0(t)
+\frac{a_0(t)}{a_1(t)}\dot a_1(t)
\end{equation}
one can get rid of $p_1$ and $p'_1$ in
eqs.(\ref{dotz},\ref{dotP}). This is due to the 
covariance of eqs.(\ref{dotz},\ref{dotP}) under the transformation
\begin{equation}
z_n(t) \rightarrow a_1(t) z_n(t)  + a_0(t),~~~~P_n(t) \rightarrow
\frac{P_n(t)}{a_1(t)}.
\end{equation}

As we are working in the gauge $\sum_n P_n=0$ it is useful to perform
the canonical transformation generated by 
\begin{equation}
G(z,P')= (z_1+\cdots + z_{\cal N}) P'_1 + (z_2- z_1)P'_2+ \dots +
(z_{\cal N} - z_1)P'_n
\end{equation}
i.e. the change of variables
\begin{equation}
z'_1 = z_1+ \dots+ z_{\cal N}
\end{equation}
\begin{equation}
z'_2 = z_2 - z_1
\end{equation}
\begin{equation}
z'_{\cal N} = z_{\cal N} - z_1
\end{equation}
\begin{equation}
P'_1 = \frac{P_1 + \dots P_{\cal N}}{\cal N}
\end{equation}
\begin{equation}
P'_n = P_n -\frac{P_1 +  \dots + P_{\cal N }}{\cal N},~~~~n>1.
\end{equation}
The reduced hamiltonian will be translational invariant,
i.e. independent of $z'_1$ to be
consistent with $\sum P_n =0$ and our canonical variables will be
$z'_2, \dots z'_{\cal N}$ and $P'_2,\dots P'_{\cal N}$.

Using the definition of ${\cal P}(z)$
\begin{equation}
\frac{1}{4\pi}
\sum_n\frac{P_{nz}}{z-z_n}=\frac{\prod_B(z-z_B)}{{\cal P}(z)} 
\end{equation}
and the properties of the locations $z_B$ of the apparent singularities
\begin{equation}\label{zB} 
\sum_n\frac{P_{nz}}{z_B-z_n}=0,
\end{equation}
one easily derives
\begin{equation} 
4 \pi \frac{\partial z'_B}{\partial P'_n} = (\frac{1}{z'_B-z'_n}
-\frac{1}{z'_B}) \frac{{\cal P}(z'_B + z_1)}{\prod_{C\neq B}(z'_B -z'_C)} 
\end{equation}
\begin{equation} 
4 \pi \frac{\partial z'_B}{\partial z'_n} = - \frac{P'_n}{(z'_B-z'_n)^2}
\frac{{\cal P}(z'_B + z_1)}{\prod_{C\neq B}(z'_B -z'_C)} 
\end{equation}
from which
\begin{equation}\label{dotz1}
\dot z'_n = -\sum_B\frac{\partial\beta_B}{\partial\mu}\frac{\partial
z'_B}{\partial P'_n} -c_1(t) z'_n~~~~~~~~n=2,\dots N 
\end{equation}
and
\begin{equation}\label{dotP1}
\dot P'_n = \frac{\partial\beta_n}{\partial\mu}+
\sum_B\frac{\partial\beta_B}{\partial\mu}\frac{\partial z'_B}{\partial
z'_n} + c_1(t)P'_n ~~~~~~~~n=2,\dots N. 
\end{equation}
Again by means of a canonical transformation of generator 
\begin{equation}
G(z',P'')= \sum_na_1(t) z'_n P''_n
\end{equation}
one can get rid of $c_1(t)$ in eq.(\ref{dotz1},\ref{dotP1}). This
holds when 
$c_1$ is simply a function of $t$. If one wants to write
eq.(\ref{dotz1},\ref{dotP1}) in the frame which does not rotate at
infinity, $c_1$ has to be chosen 
\begin{equation}\label{c-1}
c_1 = -\frac{1}{\sum_n P_n z_n}
\end{equation}
which is not simply a function of $t$. At any rate it is immediately
seen that if $H$ generates eqs.(\ref{dotz1},\ref{dotP1}) with $c_1=0$ the
hamiltonian $H+\ln(\sum_n P_n z_n \sum_n \bar P_n \bar z_n)$ generates
eqs.(\ref{dotz1},\ref{dotP1}) with $c_1$ given by eq.(\ref{c-1}).
Thus we shall here examine the case $c_1=0$.

It is instructive to treat first the three body case: Since 
there is only one apparent singularity,
the equations of motion (\ref{dotz1},\ref{dotP1}) become
\begin{equation}\label{dotz3}
\dot z'_n = -\frac{\partial\beta_A}{\partial\mu}\frac{\partial
z'_A}{\partial P'_n} %-c_1(t) z'_n 
\end{equation}
and
\begin{equation}\label{dotP3}
\dot P'_n = \frac{\partial\beta_n}{\partial\mu}+
\frac{\partial\beta_A}{\partial\mu}\frac{\partial z'_A}{\partial
z'_n}. %+ c_1(t)P'_n  
\end{equation}
 From
eq.(\ref{dotz3}) we see that the hamiltonian must be of the form
\begin{equation}\label{threehamiltonian}
H(z'_2,z'_3, P'_2,P'_3) = -\int_{z_0}^{z'_A}\frac{\partial
\beta_A}{\partial \mu}(z'_2,z'_3,z''_A) ~d z''_A + f(z'_2, z'_3)
\end{equation} 
where $z'_A$ is a function of $z'_n$ and $P'_n$ through the relation
eq.(\ref{zB}). 
With regard to the integral in the above equation, it can be
related to the $\beta_{Ar}$ appearing in the reduced $SL(2C)$
canonical equation, i.e. with $z_1\equiv 0$, $z_2\equiv 1$ recalling
that 
\begin{equation}
\beta_{A}(z'_2,z'_3,z'_A,\mu) = \frac{1}{z'_2}
\beta_A(1,\frac{z'_3}{z'_2},
\frac{z'_A}{z'_2},\mu) \equiv \frac{1}{z'_2}\beta_{Ar}(u,v,\mu)
\end{equation}
and thus 
\begin{equation}
-\int_{z_0}^{z'_A}\frac{\partial \beta_A}{\partial \mu}(z'_2,z'_3,z''_A,\mu)
~d z''_A = -\int_{z_0/z'_2}^v \frac{\partial \beta_{Ar}}{\partial
\mu}(u, v',\mu) ~d v' 
\end{equation}
with $u = z'_3/z'_2$ and $v = z'_A/z'_2$.

We must now check that with a proper choice of
$f(z'_2,z'_3)$ the hamiltonian, which already generates eqs.(\ref{dotz3}),
generates also eqs.(\ref{dotP3}). 
We have
\begin{equation}
-\frac{\partial H}{\partial z'_n} = 
\int_{z_0}^{z'_A} \frac{\partial^2
\beta_A}{\partial\mu \partial z'_n}~ d z''_A + 
\frac{\partial \beta_A}{\partial
\mu}\frac{\partial z'_A}{\partial z'_n} - \frac{\partial
f(z'_2,z'_3)}{\partial z'_n}.
\end{equation}
In appendix 1 it is proved that
\begin{equation}\label{dbAdzm}
\frac{\partial^2 \beta_A}{\partial\mu \partial z'_n} =
\frac{\partial^2 \beta_n}{\partial\mu \partial z'_A}, 
\end{equation}
from which
\begin{equation}
-\frac{\partial H }{\partial z'_n} = \frac{\partial\beta_n }{\partial
\mu} (z'_2, z'_3, z'_A,\mu) - \frac{\partial\beta_n }{\partial \mu}
(z'_2, z'_3, z_0,\mu) +\frac{\partial \beta_A}{\partial \mu}
\frac{\partial z'_A}{\partial z'_n} (z'_2, z'_3, z'_A,\mu) - \frac{\partial
f(z'_2,z'_3)}{\partial z'_n}  
\end{equation}
and thus $f(z'_2,z'_3)$ has to satisfy 
\begin{equation}
\frac{\partial f}{\partial z'_n} = -\frac{\partial\beta_n }{\partial\mu}.
\end{equation}
The integrability in $f$ of such a relation is provided by 
\begin{equation}\label{dbmdzn}
\frac{\partial\beta_n }{\partial z'_m} = \frac{\partial\beta_m }
{\partial z'_n}
\end{equation}
which is also proved in appendix 1.

\noindent
We come now to the ${\cal N}$ particle case. 
The natural extension of the three particle hamiltonian
(\ref{threehamiltonian}) is 
\begin{equation}\label{HNpart}
H(z'_2, \dots z'_{\cal N}, P'_2, \dots P_{\cal N}) =
-\int_{\{z_0\}}^{\{z'_B\}}\sum_B\frac{\partial 
\beta_B}{\partial \mu}(z'_2, \dots z'_{\cal N},z''_A, \dots,\mu) ~d z''_B +
f(z'_2, \dots z'_{\cal N}).
\end{equation}
In order eq. (\ref{HNpart}) to make sense we need that the integral be
independent of the path in the ${\cal N} - 2$ dimensional space of the
$z_B$, namely the  form $\omega =\displaystyle{\sum_A  
\frac{\partial\beta_A}{\partial\mu}dz_A}$ is exact.
Such a property is a  consequence of a conjecture due to
Polyakov \cite{ZT1} to which now we turn. Such a property
states that  the accessory parameters in the fuchsian differential
equation which solves the Liouville equation are obtained as
derivatives of the 
regularized Liouville action \cite{takh}   
$$
S_\epsilon [\phi] =\frac{i}{2} \int_{X_\epsilon} (\partial_z\phi 
\partial_{\bar z} \phi +\frac{e^\phi}{2}) dz\wedge d\bar z
-\frac{i}{2}\sum_n(1-\mu_n)\oint_n\phi(\frac{d\bar z}{\bar z -\bar
z_n}- \frac{d z}{ z - z_n})
$$
$$
+\frac{i}{2}\sum_B \oint_B\phi(\frac{d\bar z}{\bar z -\bar
z_B}- \frac{d z}{ z - z_B})
-\frac{i}{2}(\mu-2)\oint_\infty\phi(\frac{d\bar z}{\bar z}- \frac{d
z}{z})
$$
\begin{equation}\label{Sepsilon}
-\pi\sum_n(1-\mu_n)^2 \ln\epsilon^2 -\pi\sum_B\ln\epsilon^2 -\pi
(\mu-2)^2\ln\epsilon^2 
\end{equation}
computed on the solution of the Liouville equation. In
(\ref{Sepsilon}) $i dz\wedge d\bar z/2 = dx dy$ and $X_\epsilon$ is a
large disk of radius $1/\epsilon$ from which small 
disks of radius $\epsilon$ around the particles and apparent
singularities have been removed. The line integrals are all taken 
counterclockwise and they impose the correct behavior on $\phi$
around the singularities and at infinity. Polyakov conjecture states
that
\begin{equation}\label{Sepsilonmu}
-\frac{1}{2\pi} d S_\epsilon = \sum_n\beta_n dz_n + \sum_B \beta_B dz_B.  
\end{equation}
In other words, the accessory parameters $\beta_n$ and $\beta_B$
which provide $SU(1,1)$ monodromies i.e. a monodromic
conformal factor, define an exact 1-form. Such a conjecture has been
proved by 
Zograf and Tahktajan \cite{ZT1} for fuchsian differential equations
with parabolic singularities; in addition they remark that the proof
can be extended in a straightforward way to the case of elliptic
singularities of finite order. We are obviously interested in the
generic elliptic case 
including non algebraic singularities (any real $\mu_l$ with
$0<\mu_l<1$ not necessarily of the form $1/n$). The extension of 
the proof to this case seems not as straightforward since the main
tool of the proof, i.e. the mapping of 
the upper complex half plane into the punctured Riemann surface
through a properly discontinuous group, is not available. Nevertheless
from what follows it appears that such an extension is of great
relevance for the hamiltonian structure of $2+1$ gravity.

\noindent
Thus the hamiltonian 
\begin{equation}\label{Smu}
H = \frac{1}{2\pi}\frac{\partial S_\epsilon}{\partial \mu}\
\end{equation}
already provides the correct expression for $\dot
z'_n$. It is now straightforward to prove that with (\ref{Smu}) also the
equations for $\dot P_n$ are satisfied. In fact we have
\begin{equation}
-\frac{\partial H}{\partial z'_n} = -\frac{1}{2\pi}\frac{\partial^2
S_\epsilon}{\partial\mu\partial z'_n} -\frac{1}{2\pi}\sum_B\frac{\partial^2
S_\epsilon}{\partial\mu\partial z'_B} \frac{\partial
z'_B}{\partial z'_n} = \frac{\partial
\beta_n}{\partial
\mu}+\sum_B\frac{\partial\beta_B}{\partial\mu}
\frac{\partial z'_B}{\partial z'_n}.  
\end{equation}
We recall that in the non rotating frame the hamiltonian
contains an additional contribution, as already observed at the
beginning of this section. Its complete form in that 
frame is indeed given by
\begin{equation}
H= \ln\left[(\sum_n P_n z_n) (\sum_n \bar P_n \bar z_n)\right] +
\frac{1}{2\pi}\frac{\partial S_\epsilon}{\partial \mu}. 
\end{equation} 
Note that this hamiltonian, being time-indepedent,  provides a further
conservation law in the ${\cal N}-$particle problem.

\noindent
On a more formal grounds, it is interesting to notice  
that the closedness of the 1-form 
(\ref{Sepsilonmu}) is implied by the weaker relation 
\begin{equation}\label{symmAB}
\frac{\partial \beta_A}{\partial z'_B} = \frac{\partial
\beta_B}{\partial z'_A} 
\end{equation}
for the auxiliary parameters, as from these it follows through the
Garnier equations 
\begin{equation}\label{symmnA}
\frac{\partial \beta_n}{\partial z'_A} = \frac{\partial
\beta_A}{\partial z'_n}. 
\end{equation}
In fact from the Garnier equations we have
\begin{equation}
\frac{\partial \beta_A}{\partial z'_n}=-2 \frac{\partial H_n}{\partial
z'_A} - 2 \sum_B\frac{\partial \beta_A}{\partial z'_B}\frac{\partial
z'_B}{\partial z'_n} = -2 \frac{\partial H_n}{\partial
z'_A} - 2\sum_B\frac{\partial \beta_A}{\partial z'_B}\frac{\partial
H_n}{\partial \beta_B}  
\end{equation}
while
\begin{equation}
\frac{\partial \beta_n}{\partial z'_A}=-2 \frac{\partial H_n}{\partial
z'_A} - 2 \sum_B\frac{\partial H_n}{\partial \beta_B}\frac{\partial
\beta_B}{\partial z'_A} 
\end{equation}
and thus from eq.(\ref{symmAB}) we obtain eq.(\ref{symmnA}).

\noindent
Similarly we have
$$
\frac{\partial \beta_n}{\partial z'_m}=-2 \frac{\partial H_n}{\partial
z'_m} - 2 \sum_B\frac{\partial H_n}{\partial \beta_B}\frac{\partial
\beta_B}{\partial z'_m}= -2 \frac{\partial H_n}{\partial
z'_m} + 4 \sum_B\frac{\partial H_m}{\partial z'_B}\frac{\partial
H_n}{\partial \beta_B}-
$$
\begin{equation}
-4 \sum_B\frac{\partial H_n}{\partial \beta_B}\sum_C\frac{\partial
H_m}{\partial \beta_C}\frac{\partial \beta_B}{\partial z'_C}
\end{equation}
i.e. if eq.(\ref{symmAB}) holds we have
\begin{equation}
\frac{\partial \beta_n}{\partial z'_m}-\frac{\partial
\beta_m}{\partial z'_n}  = 2 \left[\frac{\partial H_m}{\partial
z'_n} -\frac{\partial H_n}{\partial z'_m}+
2 \sum_B \left(\frac{\partial H_m}{\partial
z'_B} \frac{\partial H_n}{\partial
\beta_B}-\frac{\partial H_n}{\partial
z'_B}\frac{\partial H_m}{\partial\beta_B}\right)\right]
\end{equation}
which due to a general identity \cite{okamoto} vanishes.
Thus 
\begin{equation}
\frac{\partial \beta_n}{\partial z'_m}=\frac{\partial
\beta_m}{\partial z'_n}. 
\end{equation}
Finally we notice that  eq.(\ref{Smu}) assures that the hamiltonian
$H$ is globally defined, while the integrability condition we proved
in the three particle case assures only the local existence of the
hamiltonian.   

\section{The asymptotic metric}

In the previous section we constructed the reduced particle
hamiltonian from the equations of motion. On the other hand one could
follow a different path, i.e. to recover the hamiltonian as a boundary
term in the gravitational action. In order to do so we shall first
investigate the diffeomorphism which connects the 
metric for a spinning particle in  the DJH gauge and the same geometry
in the $K=0$ gauge. It 
turns out that such a diffeomorphism can be computed exactly and that
will also allow us to compute the expression of the Killing vectors of
the spinning cone geometry in our coordinates.

The DJH metric is given by

\begin{equation}\label{DJHmetric}
ds^2 = -(dT+Jd\phi)^2 + dR^2 + \alpha^2 R^2 d\phi^2.
\end{equation}

With the transformation $R = r_0 \zeta^\alpha$ it can be put into
conformal form 
\begin{equation}
ds^2 = -(dT+Jd\phi)^2 + \alpha^2 r_0^2 (\zeta^{\alpha-1})^2 (d\zeta^2 +
\zeta^2 d\phi^2). 
\end{equation}
It is a solution of the sourceless 2+1 Einstein's equations
with a single source located at $r=0$, $\forall t$. It possesses two
Killing vector fields, ${\displaystyle\frac{\partial}{\partial T}}$ 
and ${\displaystyle\frac{\partial}{\partial \phi}}$.

For the metric in the maximally slicing gauge we shall use the ADM
form
\begin{equation}\label{ADMmetric2}
ds^2= -N^2 dt^2+ e^{2\sigma}(dz+ N^z dt)(d\bar z+ N^{\bar z} dt).
\end{equation}
We shall set with $r=|z|$
\begin{equation}
e^{2\sigma}=f^2(r,t);~~~~N^z = z n(r,t);~~~~N^{\bar z}=\bar z \bar n(r,t).
\end{equation}
Such a metric possesses the Killing vector field 
$\displaystyle{\frac{\partial} {\partial \theta}}$. Moreover
$$
g_{tt}= -N^2 + r^2 n\bar n f^2;~~~~g_{t\theta}= r^2 f^2 \frac{n -
\bar n}{2i} ;~~~g_{tr} = r f^2 \frac{n+\bar n}{2};
$$
\begin{equation}
g_{rr} =
f^2;~~~~g_{\theta\theta} = r^2 f^2;~~~~g_{r\theta}= 0. 
\end{equation}
A solution of the Einstein equations which comply to the York
instantaneous gauge is provided by
\begin{equation}
e^{2\sigma} = 2 \pi^{\bar z}_{~z} \pi^{z}_{~\bar z} e^{2\tilde\sigma}
\end{equation}
with
\begin{equation}\label{sigmatilde}
e^{-2\tilde\sigma} = \frac{8 \alpha^2}{\Lambda^2} \frac{(\frac{z\bar
z}{\Lambda^2})^{-\alpha-1}}{(1- (\frac{z\bar z}{\Lambda^2})^{-\alpha})^2}
\end{equation}
where $\alpha = 1- \mu  = 1- \frac{M}{4\pi}$.
$\tilde\sigma$ solves
\begin{equation}\label{tildesigmaeq}
\Delta (2\tilde\sigma) = - e^{-2\tilde \sigma}
\end{equation}
for $z \neq 0$ and $\pi^{\bar z}_{~z}$ is given by
\begin{equation}\label{piasympt}
\pi^{\bar z}_{~z} = -\frac{1}{2\pi z^2}\sum_n P_n z_n \equiv\frac{p(t)}{z^2}.
\end{equation}
Eq.(\ref{piasympt}) is the asymptotic form of the expression 
\begin{equation}
-\frac{1}{2\pi} \sum_n\frac{P_n}{z-z_n}
\end{equation}
subject to the constraint $\sum_n P_n =0$. 

$N$ and $N^z$ are given by eq.(\ref{N},\ref{Nz}).
We know that $p(t)$ evolves according to
\begin{equation}
p(t) = -\frac{1}{2\pi}[\alpha t -ib]
\end{equation}
and $\Lambda$ is given by
\begin{equation}\label{Lambda}
\Lambda^2(t) = c_\Lambda [p(t) \bar p(t)]^{\frac{1}{\alpha}}.
\end{equation}
In fact if the conformal factor $e^{2\sigma}$ has to provide a
solution of Einstein's equations the coefficient $s^2$ which appears in
its asymptotic expansion
\begin{equation}
e^{2\sigma} \approx s^2 (z\bar z)^{-\mu}
\end{equation}
has to be time independent. We shall  see in the following section
that $\ln s^2$ coincides with the hamiltonian, which is obviously
conserved.  
One checks that the metric (\ref{ADMmetric2}) with positions
(\ref{N},\ref{Nz},\ref{sigmatilde},\ref{piasympt},\ref{Lambda})  
satisfy Einstein's 
equation in all space with a source confined to $z=0$.

In order to find the diffeomorphism which connects the two metrics
(\ref{DJHmetric})
and (\ref{ADMmetric2}) it is useful to introduce the intermediate
variable $\rho = 
(\frac{r}{\Lambda})^\alpha$. We have
\begin{equation}
-2\tilde\sigma= \ln(\frac{8}{\Lambda^2})+ 2 \ln \alpha -
\frac{1-\alpha}{\alpha}\ln\rho^2 - 2 \ln(1-\rho^{-2})
\end{equation}
and
\begin{equation}
N =\frac{1}{2 \pi\alpha}\left[\frac{\ln (\rho^2 k^2)}{2} -1
+\frac{1}{\rho^2-1}\ln(\rho^2k^2)\right] 
\end{equation}
where
\begin{equation}
\ln k^2 = 2 \alpha^2 \frac{\partial \ln\Lambda^2}{\partial \mu}.
\end{equation}
$k^2$ in general will be a function of time and given a solution of
the ${\cal N}$- particle problem it is a well defined function;
e.g. it can be explicitly computed in the two body case. On the other
hand one can verify that the asymptotic metric provided by
eq.(\ref{N},\ref{Nz},\ref{sigmatilde},\ref{piasympt}) is a solution of
Einstein's equations 
for any $k^2(t)$ (see appendix 2). 
It is important to note that the solution of
\begin{equation}
\Delta N = e^{-2\tilde\sigma} N
\end{equation}
obtained by taking the derivative of eq.(\ref{tildesigmaeq})
with respect to $\mu$ is performed at fixed time and the $\Lambda$
appearing in eq.(\ref{sigmatilde})  is a function of $\mu$. From
eq.(\ref{Nz}) $N^z$ is 
given by 
\begin{equation}
N^z = \frac{z}{\pi p(t)}\left[\frac{\rho^2\ln(\rho^2 k^2)}{(\rho^2-1)^2}
-\frac{1}{(\rho^2-1)}\right] \equiv z n.
\end{equation}
The most general transformation which transforms
$\displaystyle{\frac{\partial}{\partial \theta}}$ into 
$\displaystyle{\frac{\partial}{\partial\phi}}$ is
\begin{equation}
R = R(\rho,t);~~~~ T= T(\rho,t);~~~~\phi = \theta +\omega(\rho,t).
\end{equation}
Equating the coefficient of $d\theta^2$ one obtains with $f^2 = e^{2\sigma}$
\begin{equation}\label{gthth1}
g_{\theta\theta} = r^2 f^2 = \alpha^2 R^2 - J^2
\end{equation}
and we have for the metric in the variables $t,\theta,\rho$
\begin{equation}
g_{\rho\rho} = \frac{|p(t)|^2}{4\alpha^4}(1 - \rho^{-2})^{2}
\end{equation}
and
\begin{equation}\label{gthth2}
g_{\theta\theta} = \frac{|p(t)|^2}{4 \alpha^2}(\rho - \rho^{-1})^{2}.
\end{equation}
 From eqs.(\ref{gthth1},\ref{gthth2}) we obtain
\begin{equation}\label{R2}
R^2 = \frac{1}{\alpha^2}(J^2+\frac{|p(t)|^2}{4\alpha^2} 
(\rho - \rho^{-1})^2).
\end{equation}
The matching of the $\rho\theta$ and $\rho\rho$ components of the
metric gives
\begin{equation}\label{thetarho}
0=g_{\theta\rho} = -J(\partial_\rho T + J\partial_\rho\omega) +\alpha^2
R^2\partial_\rho\omega 
\end{equation}
\begin{equation}\label{rhorho}
g_{\rho\rho} = -(\partial_\rho T+J\partial_\rho\omega)^2 +(\partial
R)^2 +\alpha^2 R^2(\partial_\rho\omega)^2
\end{equation}
from which we deduce
\begin{equation}\label{domega2}
(\partial_\rho\omega)^2 = \frac{J^2[(\partial_\rho R)^2 -
g_{\rho\rho}]}{\alpha^2R^2g_{\theta\theta}}
\end{equation}
and substituting eqs.(\ref{gthth1},\ref{R2},\ref{rhorho}) into
(\ref{domega2}) we have 
\begin{equation}
\partial_\rho\omega = \frac{4 J\rho
V(t)}{|p(t)|^2(\rho^2-1)+4\alpha^2J^2\rho^2} = \frac{2\rho B
}{\alpha[(\rho^2-A)^2 +B^2]} 
\end{equation}
with $V(t) =  \sqrt{|p(t)|^2 - \alpha^2 J^2}$, $\displaystyle{A(t)=
1-\frac{2\alpha^2 J^2}{|p(t)|^2}}$, $\displaystyle{B(t)= \frac{2\alpha
J V(t)}{|p(t)|^2}}$,
which integrated gives
\begin{equation}
\omega = \frac{1}{\alpha}\left[\arctan\left(\frac{\rho^2-A}{B}\right)
-\frac{\pi}{2}\right] + f(t) \equiv \bar\omega(t,\rho) + f(t).
\end{equation}
$\bar\omega$ for $\rho\rightarrow\infty$ goes to zero.

Similarly from eqs.(\ref{thetarho},\ref{rhorho}) we find
\begin{equation}
\partial_\rho T =
\frac{V(t)}{\alpha^2 \rho}- J\partial_\rho \omega
\end{equation}
from which
\begin{equation}
T= \frac{V(t)}{2\alpha^2}\ln\rho^2 
- J\bar\omega(t,\rho) +h(t) \equiv \bar T(t,\rho) + h(t).
\end{equation}
The matching of the $t \theta$ component of the metric gives
\begin{equation}
g_{t\theta} = -J\partial_t T + (\alpha^2 R^2 - J^2) \partial_t\omega
\end{equation}
i.e.
$$
-\frac{b}{8\pi^2 \alpha^2}\left[ \ln(k^2\rho^2)-1+\rho^{-2}\right]=
$$
\begin{equation}
- J \dot h(t) - \frac{J\dot V}{2\alpha^2}\ln\rho^2 +
\frac{|p(t)|^2}{4\alpha^3}[-\dot B + (A\dot B- \dot A B)\rho^{-2}] +
\alpha^2 R^2 \dot f(t).   
\end{equation}
As $R^2(\rho)$ behaves like $\rho^2$ for large $\rho$, we have $\dot
f(t)=0$ which means that the two frames asymptotically do not rotate
one with respect to the other. The $\ln\rho^2$ terms fixes $b = 2\pi
\alpha J$ thus giving $V(t) = \frac{\alpha t}{2\pi}$,
while the matching of the constant terms gives
\begin{equation}\label{doth}
\dot h(t) = \frac{1}{4\pi\alpha}\left[ \ln k^2(t) - \frac{2\alpha^2
J^2}{|p(t)|^2} \right],
\end{equation}
which defines $h(t)$ up to a constant; this is due to the fact that in
the DJH gauge the time like Killing vector is simply
$\frac{\partial}{\partial T}$.   
Now the diffeomorphism is completely fixed and one can check that the
remaining equations for $g_{tt}$ and $g_{t\rho}$ are satisfied.

Summarizing the diffeomorphism is given by 
\begin{equation}\label{Req} 
R^2 =
\frac{1}{\alpha^2} \left[J^2+ \frac{r^{2\alpha}}{4 c_\Lambda \alpha^2}
\left( 1 - c_\Lambda |p(t)|^2 r^{-2\alpha}\right)^2\right] 
\end{equation}
\begin{equation} \label{phieq}
\phi = \theta +\omega \equiv\theta+ \frac{1}{\alpha}\arctan\left[2\pi
\frac{c^{-1}_\Lambda r^{2\alpha} - 
|p(t)|^2 + 2 \alpha^2 J^2}{2 \alpha^2 J t} \right]
\end{equation}
\begin{equation}\label{Teq}
T = \frac{t}{4\pi} \left[\ln \frac{r^2}{c_\Lambda}
-\frac{1}{\alpha}\ln|p(t)|^2\right] - J \omega +h(t) 
\end{equation} 
with $h(t)$ obeying eq.(\ref{doth}).
This shows that two asymptotic solutions with different $k^2(t)$ are
diffeomorphic to the same DJH metric and thus are diffeomorphic to
each other.  This explains why for any choice of $k^2(t)$
eq.(\ref{N},\ref{Nz},\ref{sigmatilde}) are solutions of Einstein's
equations.  
For large $r$ eqs.(\ref{Req},\ref{phieq},\ref{Teq}) become
\begin{equation} 
R^2 \approx
\frac{r^{2\alpha}}{4c_\Lambda \alpha^4}
\end{equation}
\begin{equation} 
\phi \approx \theta + \frac{\pi}{2\alpha}
\end{equation}
\begin{equation} 
T \approx \frac{t}{4\pi} \ln(\frac{r^2}{c_\Lambda
|p(t)|^{\frac{2}{\alpha}}}) -\frac{\pi J}{2\alpha} +h(t).
\end{equation} 
In the DJH gauge a finite
transformation along the Killing vector
$\displaystyle{\frac{\partial}{\partial T}}$ is 
simply given by $ T \rightarrow T+ c$ while in the York 
instantaneous gauge it is more complicated. The time-like Killing
vector in the instantaneous York gauge is simply computed and given by 
\begin{equation}\label{killing} 
\frac{(2\pi)^3\alpha (\rho^2+1) |p(t)|^2}{{\cal D}}
\frac{\partial}{\partial t} + \frac{8\pi^2 J \alpha^2}{{\cal
D}}\frac{\partial}{\partial \theta} +
\frac{4\pi \alpha^2 r t}{{\cal D}} \frac{\partial}{\partial r}
\end{equation} 
with
\begin{equation}
{\cal D} = 4\pi^2 |p(t)|^2 (\rho^2+1) [\ln\rho +2\pi\alpha \dot h(t)]
+\alpha^2 t^2(1-\rho^2) + 8 \pi^2 J^2 \alpha^2.  
\end{equation} 
For large $r$ the vector (\ref{killing}) reduces to
\begin{equation}
\frac{4\pi}{\ln(\frac{r^2}{c_\Lambda
|p(t)|^{\frac{2}{\alpha}}})}\left(\frac{\partial}{\partial t} +  
\frac{ \alpha J
c_\Lambda^\alpha}{\pi r^{2\alpha}}\frac{\partial}{\partial \theta} +  
\frac{ \alpha t c_\Lambda^\alpha }{2 \pi^2 r^{2\alpha}}r
\frac{\partial}{\partial r}\right). 
\end{equation} 

\section{The hamiltonian as a boundary term}

We have solved the hamiltonian and diffeomorphism constraints and
moreover in the $K=0$ conformal gauge we have $\pi^{ij}\dot
g_{ij}\equiv 0$. Thus the action of the particles plus gravity reduces to
\begin{equation}
S  = \int dt (\sum_n P_{ni}\, \dot q_n^i -H_B)
\end{equation}
with 
\begin{equation}\label{HB1}
H_B= -2\int dt \int_{B_t} d^{(D-1)}x \,\sqrt{\sigma_{Bt}} N 
\left( K_{B_t}+\frac{\eta}{\cosh\eta}{\cal D}_\alpha v^\alpha\right)
+2\int dt \int_{B_t} d^{(D-1)}x\, r_\alpha
\pi^{\alpha\beta}_{(\sigma_{Bt})} N_\beta.
\end{equation} 

We want now to extract from $H_B$ the reduced particle hamiltonian and
compare it to the hamiltonian $H$ derived directly from the particle
equations of motion.  

The last term in the above equation can be computed as follows: on the
boundary $x^2+y^2=r_0^2= {\rm const}$ we have
\begin{equation} 
r_\alpha \pi^{\alpha\beta}_{(\sigma_{Bt})} N_\beta = -2(\bar z
\partial_{\bar z} N + z \partial_{z} N) + \bar z g(\bar z) \pi^z_{~\bar
z} + z g(z) \pi^{\bar z }_{~z} 
\end{equation}
whose integral in $d\theta$ between $0$ and $2\pi$ is given by 
\begin{equation}\label{Nzt}  
-2 \oint (\bar z \partial_{\bar z} + z \partial_{z}) N d\theta +
i\oint d\bar z  g(\bar z) \pi^z_{~\bar z} -i \oint dz g(z) \pi^{\bar z
}_{~z}.
\end{equation} 
As for large $|z|$, $N$ behaves like $\ln(z\bar z)/4 \pi$ we see that
the first term in the above expression goes over to the constant
$-2$. In the computation of the remaining 
terms as already noticed in \cite{MS1} the only contribution in $g(z)$
which survives in the sum is the one arising from the linear term in
the first order polynomial $p_1(z)$ which in the frame non rotating at
infinity is given by
\begin{equation} 
p_1(z) = c_0 -~\frac{1}{\sum_n P_n z_n} ~z.
\end{equation} 
Using this result we find zero for eq.(\ref{Nzt}) i.e. for the last term
in (\ref{HB1}). 
Similarly one proves that the contribution of the term ${\cal
D}_\alpha v^\alpha$ goes to zero like $(r_0^2)^{\mu-1}\ln r_0^2$ for
$r_0\rightarrow \infty$. Thus we are left with the boundary term
\begin{equation}\label{HBasint}
H_B  = -2 \int_{B_t} d^{(D-1)}x \,\sqrt{\sigma_{Bt}} N K_{B_t}.
\end{equation} 
By inserting the metric eqs.(\ref{N},\ref{Nz},\ref{sigmatilde}) into
the expression for $K_{Bt}$ 
and $\sigma_{Bt}$ we obtain for the integral
\begin{equation}\label{H_B}
H_B = - 4\pi N r_0 \partial_r[\ln(r e^{\sigma})]
\end{equation} 
and thus for large $r_0$ the boundary term
becomes
\begin{equation}\label{HBr0}
H_B = - r_0\ln r_0^2 (\frac{1}{r_0}+ \partial_r \sigma)=
(\mu-1) \ln r_0^2. 
\end{equation} 
We recall now that the equations of motion are obtained from the
action by keeping the values of the fields fixed at the boundary, or
equivalently \cite{wald} by keeping fixed the intrinsic metric of
the boundary. In our case the variations should be performed keeping
fixed the fields $N$, $N^a$, and $\sigma$ at the boundary. %Let us
%consider a given value $\mu_0$ of $\mu$. 
We shall perform the
computation for the boundary given by a circle of radius $r_0$ for a
very large value of $r_0$.  If we change the positions of 
particle positions and momenta, $\Lambda$ varies and in order to keep
the value of $\sigma$ fixed at the boundary we must vary $\mu$ as to
satisfy the following equality
\begin{equation}
\ln\{(\sum_n P_n z_n)(\sum_n \bar P_n \bar z_n)\} - \mu \ln r_0^2
+(\mu-1)\ln\Lambda^2 -\ln 16\pi^2 
\equiv -\mu\ln r_0^2 + \ln s^2 = {\rm const.} 
\end{equation} 
Thus
\begin{equation}
0= - \delta \mu \ln r_0^2 +\sum_n( \delta z_n \frac{\partial \ln
s^2}{\partial z_n} + \delta P_n \frac{\partial \ln
s^2}{\partial P_n} + {\rm c.c.} ) + \delta \mu \frac{\partial \ln
s^2}{\partial \mu}
\end{equation} 
i.e. for large $r_0$
\begin{equation}
\delta \mu \approx \frac{1}{\ln r_0^2}\sum_n( \delta z_n \frac{\partial \ln
s^2}{\partial z_n} + \delta P_n \frac{\partial \ln
s^2}{\partial P_n} + {\rm c.c.} ).
\end{equation} 
Substituting % the obtained value of $\mu$ 
into eq.(\ref{HBr0}) %the hamiltonian becomes
we have 
\begin{equation}
\delta H_B = \sum_n( \delta z_n \frac{\partial \ln
s^2}{\partial z_n} + \delta P_n \frac{\partial \ln
s^2}{\partial P_n} + {\rm c.c.} )
\end{equation} 
i.e. apart for a constant $H_B$ equals $ \ln s^2$ 
\begin{equation}\label{Hr}
H_B = \ln s^2 +{\rm const.} = \ln\left[(\sum_n P_n z_n)(\sum_n \bar P_n
\bar z_n)\right] +(\mu-1)\ln\Lambda^2 +{\rm const.}
\end{equation}

In the two particle case one can check that eq.(\ref{Hr}) coincides
with the
hamiltonian derived directly from the equations of motion.
In fact explicit computation by using the expression of $\Lambda$ in
terms of hypergeometric functions gives
\begin{equation}
\Lambda^2 = |z'_2|^2 \left[ \frac{1}{8(1-\mu)^2
G(\mu)}\right]^{\frac{1}{1-\mu}}
\end{equation}
where
$$
G(\mu) = \pi^{-2}\Gamma^4(1-\mu)\sin^2(\pi \mu)\times
$$ 
\begin{equation}
\Delta((\mu+\mu_1+\mu_2)/2)\Delta((\mu-\mu_1+\mu_2)/2)
\Delta((\mu+\mu_1-\mu_2)/2)\Delta((\mu-\mu_1-\mu_2)/2).  
\end{equation}
The boundary term eq.(\ref{HBasint}) depends on the fields on the
boundary and also on the derivative of the fields directed towards the
interior i.e. derivative with respect to $r$. By keeping the values of
the fields fixed on the boundary it provides the hamiltonian, i.e. a
function of $z_n$ and $P_n$ which through Hamilton's equations give
rise to the equations of motion. It is not however the energy as
usually defined i.e. the value of the boundary term when $(N, N^i)$
take the values of the asymptotic time-like Killing vector. In our
case due to the choice of the $K=0$ gauge which vastly simplifies the
dynamics, the $(N, N^i)$ differ from the timelike asymptotic Killing
vector. The energy of a solution is easily obtained in the DJH gauge,
where one checks from the metric eq.(\ref{DJHmetric}) that with $(N,
N^i)= (1,0,0)$ i.e the normalized Killing vector, one obtains for $H_B$
the value $4\pi (\mu -1)$ as expected. 

It is of interest to examine how $\ln\Lambda^2$ behaves under a
complex scaling $z' = \alpha z$. It is easily seen  from the Liouville
equation that if $2\tilde\sigma(z)$ is a solution with singularities
in $z_n$ and $z_B(z_n, P_n)$ the solution with singularities in
$\alpha z_n$ and $z_B(\alpha z_n, P_n/\alpha) = \alpha z_B(z_n, P_n)$
is given by
\begin{equation}
2\tilde\sigma(z) = 2\tilde\sigma(\frac{z}{\alpha}) +\ln(\alpha\bar\alpha). 
\end{equation}
It implies the following transformation law on $\ln\Lambda$
\begin{equation}
\ln\Lambda^2(\alpha z_n, \frac{P_n}{\alpha})=\ln\Lambda^2(z_n,P_n)
+\ln(\alpha\bar\alpha)  
\end{equation}
which provides the following Poisson bracket
\begin{equation}
[H,\sum_nP_n z_n] = [\sum_nP_n z_n, (\mu-1) \ln\Lambda^2] = \mu-1 
\end{equation}
and thus we have reached a hamiltonian derivation of the generalized
conservation law
\begin{equation}
\sum_nP_n z_n = (1-\mu)(t-t_0)-ib.
\end{equation}

We want now to relate the result eq.(\ref{Hr}) to the results of
sect.4. 
Let us now consider the value of the
action $S_\epsilon$ on the solution of the Liouville
equation and let 
us compute its derivative with respect to $\mu$. As we are varying
around a stationary point the only contribution is provided by the
terms in eq.(\ref{Sepsilon}) which depend explicitly on $\mu$ i.e.
\begin{equation}
\frac{\partial S_\epsilon}{\partial \mu} = -\frac{i}{2}
\oint_\infty\phi(\frac{d\bar z}{\bar z}- \frac{d z}{ z}) - 2\pi
(\mu-2)\ln\epsilon^2  
\end{equation}
and as $\phi\equiv -2\tilde\sigma$ at infinity behaves like
\begin{equation}
\phi\approx \ln 8(1-\mu)^2 +(\mu-2)\ln z\bar z-(\mu-1)\ln\Lambda^2
\end{equation}
we have
\begin{equation}
\frac{\partial S_\epsilon}{\partial \mu} = -2\pi \ln8(1-\mu)^2 + 2\pi
(\mu-1)\ln\Lambda^2.
\end{equation}
Thus we can rewrite eq.(\ref{Hr}) as
\begin{equation}
H_r = \ln\left[(\sum_n P_n z_n) (\sum_n \bar P_n \bar z_n)\right] +
\frac{1}{2\pi}\frac{\partial S_\epsilon}{\partial \mu} + {\rm const}
\end{equation}
in agreement with the result of sect.4 obtained through Polyakov's
conjecture.  
\section{Quantization: the two particle case}

We recall that the classical two particle hamiltonian in the reference
system which does not rotate at infinity is given by
\begin{equation}
H = \ln(P z \bar P \bar z) + (\mu -1)\ln (z\bar z)=
\ln(Pz^\mu) + \ln(\bar P\bar z^\mu) = h + \bar h
\end{equation}
with $P = P'_2$ and $z = z'_2$.
$h$ and $\bar h$ are separately constant of motion and if we combine
them with the generalized conservation law $Pz = (1-\mu)(t-t_0) -ib$
(see eq.(\ref{dilatation})) we obtain the solution for the motion
\begin{equation}
z = {\rm const} [(1-\mu)(t-t_0) - ib]^{\frac{1}{1-\mu}}. 
\end{equation}
$H$ can be rewritten as
\begin{equation}
H = \ln((x^2+y^2)^\mu ((P_x)^2 + (P_y)^2)).
\end{equation}
Keeping in mind that with our definitions $P$ is the momentum
multiplied by $16\pi G_N/c^3$, applying the correspondence principle 
we have
\begin{equation}
[x,P_x] = [y,P_y] = i l_{P}
\end{equation}
where $l_P = 16 \pi G_n\hbar/c^3$,
all the other commutators equal to zero. $H$ is converted into the
operator
\begin{equation}\label{logbeltrami}
\ln[-(x^2+y^2)^\mu \Delta] +~{\rm constant}.
\end{equation}
The argument of the logarithm is the Laplace-Beltrami
$\Delta_{LB}$ operator on the metric $ds^2=(x^2+y^2)^{-\mu} (dx^2 +
dy^2)$. Following 
an argument similar to the one presented in   
\cite{sorkin} one easily proves that if we start from the 
domain of $\Delta_{LB}$ given by the infinite differentiable functions
of compact support $C^\infty_0$ which can also include the origin,
then $\Delta_{LB}$ has a unique 
self-adjoint extension in the Hilbert space of functions square
integrable on the metric $ds^2=(x^2+y^2)^{-\mu} (dx^2+dy^2)$ and as a
result since $\Delta_{LB}$ is a positive operator, $\ln(\Delta_{LB})$ is
also self-adjoint. In fact expanding in circular harmonics
\begin{equation}
\psi(x,y) = \sum_m \frac{e^{im\theta}}{\sqrt{2\pi}}\phi_m(r)
\end{equation}
the indicial equation furnishes the  behaviors $r^{\pm m}$ at the
origin for $m\neq 0$ and $r^0$, $\ln(r)$ for $m=0$. Then for $m\neq 0$
only the behavior $r^{|m|}$ gives rise to a square integrable function
(we recall that $\mu<1$). For $m=0$ if $\Delta_{LB}$ is defined already
on the $C^\infty_0$ functions with support which can include the
origin, one sees that the equation
$(\Delta_{LB}^* \pm i) \phi =0$ has no solution for $\phi\in
D(\Delta^*_{LB})$. In fact if $D(\Delta_{LB})$ includes the
$C^\infty_0$ functions whose support can include the origin, then
$D(\Delta^*_{LB})$ cannot contain functions which diverge
logarithmically at the origin. But if the $\phi$ which solves
$(\Delta_{LB}^* \pm i) \phi =0$ has no logarithmically divergent part
one proves easily that $(\phi, \Delta_{LB} \phi) = {\rm real}$, which
is a contradiction. Obviously  in rewriting $H$ in the from
eq.(\ref{logbeltrami}) a well defined ordering has been chosen; one
that appears rather appealing due the simplicity and covariant
nature of the result. 

Deser and Jackiw \cite{deserjackiw} considered the quantum scattering
of a test particle 
on a cone both at the relativistic and non relativistic level. Most of
the techniques developed there can be transferred here. The main
difference is the following; instead of the 
hamiltonian $(x^2+y^2)^\mu(p_x^2+p_x^2)$ which appears in their non
relativistic treatment, we have now the hamiltonian
$\ln[(x^2+y^2)^\mu(p_x^2+p_y^2)]$. The partial wave eigenvalue
differential equation 
\begin{equation}
(r^2)^\mu[-\frac{1}{r} \frac{\partial }{\partial r} r \frac{\partial
}{\partial r}+\frac{n^2}{r^2} ]\phi_n(r) = k^2 \phi_n(r) 
\end{equation}
with $\mu=1-\alpha$ is solved by
\begin{equation}
\phi_n(r) = J_\frac{|n|}{\alpha}(\frac{k}{\alpha}r^\alpha)
\end{equation}
and we have the completeness relation
\begin{equation}
\delta^2(z-z')=\alpha \sum_n \frac{e^{in(\phi-\phi')}}{2\pi}\int_o^\infty 
\frac{r^{(\alpha-1)}}{\alpha} J_\frac{|n|}{\alpha}(\frac{k}{\alpha}r^\alpha)
k dk \frac{{r'}^{(\alpha-1)}}{\alpha}
J_\frac{|n|}{\alpha}(\frac{k}{\alpha}{r'}^\alpha) 
\end{equation}
from which the logarithm of the operator $-\Delta_{LB}$, which is
our hamiltonian, becomes
\begin{equation}
\alpha\sum_n \frac{e^{in(\phi-\phi')}}{2\pi}\int_o^\infty 
\frac{r^{(\alpha-1)}}{\alpha} J_\frac{|n|}{\alpha}(\frac{k}{\alpha}r^\alpha)
\ln(k^2) k dk \frac{{r'}^{(\alpha-1)}}{\alpha}
J_\frac{|n|}{\alpha}(\frac{k}{\alpha}{r'}^\alpha).
\end{equation}
It is a self-adjoint operator with domain \cite{riesznagy} given by
those $f(z)$ such that
\begin{equation}
\alpha \sum_n \int(\ln(k^2))^2 k dk
|\int_o^\infty J_\frac{|n|}{\alpha}(\frac{k}{\alpha}r^\alpha)
\frac{r^{\alpha-1}}{\alpha}f_n(r) r dr |^2 < \infty.
\end{equation}
The Green function is given by
\begin{equation}
G(z,z',t) = \alpha \sum_n \frac{e^{in(\phi-\phi')}}{2\pi}\int_o^\infty 
\frac{r^{(\alpha-1)}}{\alpha} J_\frac{|n|}{\alpha}(\frac{k}{\alpha}r^\alpha)
(k^2)^{-i c t/l_P} k dk \frac{{r'}^{(\alpha-1)}}{\alpha}
J_\frac{|n|}{\alpha}(\frac{k}{\alpha}{r'}^\alpha).
\end{equation}
The integral in $k$ can be performed in terms of hypergeometric
functions, to obtain
$$
G(z,z',t) =
\frac{2}{ \alpha\Gamma(\frac{ict}{l_P})r r'} \left (\frac{r^\alpha
+{r'}^\alpha}{2\alpha}\right )^{2 
ict/l_P} 
$$ 
\begin{equation}
\sum_n \frac{e^{in(\phi-\phi')}}{2\pi}
\frac{\Gamma(\frac{|n|}{\alpha}+1 -
\frac{i c t}{l_P})}{\Gamma(\frac{|n|}{\alpha}+1)} 
\rho^{\frac{|n|}{\alpha}+1}{_2F_1}(\frac{|n|}{\alpha}+1 - \frac{ict}{l_P};
\frac{|n|}{\alpha}+\frac{1}{2}; 2 \frac{|n|}{\alpha}+1; 4 \rho)
\end{equation}
where
\begin{equation}
\rho = \frac{r^\alpha {r'}^\alpha}{r^\alpha+{r'}^\alpha}.
\end{equation}

\section*{Acknowledgments}

We are grateful to Marcello Ciafaloni and Stanley Deser for
interesting discussions.

\section*{Appendix 1: Properties of the residues $\beta_n$, $\beta_A$
in the three body problem}

In this appendix we outline the derivation of relations
(\ref{dbAdzm},\ref{dbmdzn}) which were used in sect.4 to prove the
hamiltonian nature of the equations for $\dot z'_n$ and $\dot P'_n$ of
the three body problem.  

To this end we shall exploit the Garnier
equations which express the isomonodromic evolution of the apparent
singularity \cite{yoshida}. We recall that the Garnier equations are a
direct outcome 
of the ADM treatment of the particle dynamics \cite{MS1}. The Garnier
hamiltonian 
for the standard parameters $z_A$ and $b_A$ ($z_1\equiv 0$, $z_2\equiv
1$ see \cite{yoshida,okamoto}) is given by
\begin{equation}\label{garnierhamiltonian}
H_G=\frac{z_A(z_A-1)(z_A - z_3)}{z_3(z_3-1)}\left\{b_A^2 - (\frac{\mu _1}{z_A}
+\frac{\mu _2} {z_A-1} +\frac{\mu _3-1}{z_A-z_3})b_A +
\frac{\kappa}{z_A(z_A-1)} \right\}
\end{equation}
and we have
\begin{equation}
\frac{\partial z_A}{\partial z_3}  = \frac{\partial H_G}{\partial
b_A};~~~~ \frac{\partial b_A}{\partial z_3}  = -\frac{\partial
H_G}{\partial z_A}.  
\end{equation}
$b_A$ is related to the $\beta_{A}$ appearing in the $SL(2C)$
canonical form (again $z_1\equiv 0$, $z_2\equiv 1$) by
\begin{equation}
b_A =\frac{\beta_{A}}{2} -\frac{1}{2}\left( \frac{1-\mu _1}{z_A}
+\frac{1-\mu _2}{z_A-1}+ \frac{1-\mu _3}{z_A-z_3}\right). 
\end{equation}
Starting from the previous equations it is not difficult to verify
directly (see also \cite{okamoto}) that the two Garnier hamiltonians
$H_2$ and $H_3$ which supervise the evolution of the auxiliary
parameters $z'_A$ and $\beta_A \equiv 2 \tilde\beta_A$ appearing in
the $SL(2C)$ 
canonical differential equation 
\begin{equation}
y''(z)+ Q(z) y(z) =0
\end{equation}
with
$$
Q(z) =  \frac{1-\mu^2_1}{4 z^2} +\frac{1-\mu^2_2}{4(z-z'_2)^2}+
\frac{1-\mu^2_3}{4(z-z'_3)^2} - \frac{3}{4(z-z'_A)^2} 
$$
\begin{equation}
+\frac{\beta_1}{2 z} +\frac{\beta_2}{2(z-z'_2)} +
\frac{\beta_3}{2(z-z'_3)} + \frac{\beta_A}{2(z-z'_A)},  
\end{equation}
are given by $H_2 = -\beta_2/2$ and $H_3 = -\beta_3/2$, i.e. by the
simple residues at $z'_2$ and $z'_3$ of the function $Q(z)$.
 This is
obtained by expressing $\beta_2$ and $\beta_3$ in terms of $z'_2, z'_3,
z'_A, \beta_A, \mu $ using the Fuchs and no-logarithm conditions at $z'_A$. 
Under isomonodromic transformations generated by a change
in the value of $z'_n$ we have
\begin{equation}
\frac{d z'_A}{d z'_n}  = \frac{\partial
H_n}{\partial \tilde\beta_A};~~~~ 
\frac{d \tilde\beta_A}{d z'_n}  = -\frac{\partial
H_n}{\partial z'_A}.
\end{equation}
The monodromy condition on $\tilde\sigma$ or equivalently the
imposition of the $SU(1,1)$ nature of the monodromies described by the
fuchsian differential equation fixes $\beta_A$ and consequently,
through Fuchs and no logarithm conditions,
$\beta_1, \beta_2, \beta_3$ as functions of $z'_1, z'_3, z'_A, \mu$, 
i.e.
\begin{equation}
H_n(z'_2, z'_3, z'_A,\mu, \beta_A(z'_2,z'_3,z'_A,\mu)) =
- \frac{1}{2}\beta_n(z'_2, z'_3, z'_A,\mu).
\end{equation}
Consider now that due to the Garnier equations we have
\begin{equation}
\frac{\partial \beta_A}{\partial z'_n}= -2 \frac{\partial
H_n}{\partial z'_A} - 2 \frac{\partial \beta_A}{\partial
z'_A}\frac{\partial H_n}{\partial \beta_A} 
\end{equation}
and also
\begin{equation}
\frac{\partial \beta_n}{\partial z'_A}= -2 \frac{\partial
H_n}{\partial z'_A} - 2 \frac{\partial H_n}{\partial
\beta_A}\frac{\partial \beta_A}{\partial z'_A}=\frac{\partial
\beta_A}{\partial z'_n} 
\end{equation}
which proves eq.(\ref{dbAdzm}) of the text. 

Similarly we have
$$
\frac{\partial \beta_n}{\partial z'_m}=-2 \frac{\partial H_n}{\partial
z'_m} - 2 \frac{\partial H_n}{\partial \beta_A}\frac{\partial
\beta_A}{\partial z'_m}= -2 \frac{\partial H_n}{\partial
z'_m} + 4 \frac{\partial H_m}{\partial z'_A}\frac{\partial
H_n}{\partial \beta_A}-
$$
\begin{equation}
-4 \frac{\partial H_n}{\partial \beta_A}\frac{\partial
H_m}{\partial \beta_A}\frac{\partial \beta_A}{\partial z'_A}
\end{equation}
and thus
\begin{equation}\label{appdbmdbn}
\frac{\partial \beta_n}{\partial z'_m}-\frac{\partial
\beta_m}{\partial z'_n}= 2 [ \frac{\partial H_m}{\partial
z'_n} - \frac{\partial H_n}{\partial z'_m} +
2 (\frac{\partial H_m}{\partial z'_A}  \frac{\partial H_n}{\partial
\beta_A} -\frac{\partial H_n}{\partial z'_A} \frac{\partial H_m}
{\partial\beta_A})]. 
\end{equation}

Using the expression of $H_2$ and $H_3$ in terms of $z'_n$, $z'_A$,
$\beta_A$ and $\mu$, obtained from the Fuchs and no logarithm
relations, one can check directly that the r.h.s. of
eq.(\ref{appdbmdbn}) is zero thus proving eq.(\ref{dbmdzn}) of the
text.

\section*{Appendix 2: General solution for the N with spherical symmetry}
In the following we shall investigate the structure of the most general 
spherically symmetric solution of the equation
\begin{equation}
\label{eqN}
\triangle N= N \exp(-2\tilde\sigma),
\end{equation}
where the conformal factor is given by  (\ref{sigmatilde}).  In polar
coordinates,  
where the requirement of spherical symmetry becomes
$\displaystyle{\frac{\partial N}{\partial\theta}=0}$, the above
equation takes the form 
\begin{equation}
\label{eqN1}
r\frac{\partial }{\partial r}\left ( r \frac{\partial N }{\partial r}\right)
= r^2 \exp(-2\tilde\sigma(t,r)) N.
\end{equation} 
If we set $r=\exp(w-w_0)$, it reduces to the following ordinary 
linear differential equation of the second order
\begin{equation}
\label{eqN3}
\frac{\partial^2 N}{\partial w^2}= \exp(-2\tilde\sigma(t,w)+2(w-w_0)) N.
\end{equation}
Given a particular solution $\bar N$ of eq.(\ref{eqN3}), a second one
can be sought in the factorized form $\phi(w,t) \bar N (w,t)$. A
short calculation shows that $\phi(w,t)$ must satisfy 
\begin{equation}
\frac{\partial}{\partial w}\left ( \log \left (\frac{\partial
\phi}{\partial w} 
\right )\right)=-\frac{\partial}{\partial w} \log \bar N^2,
\end{equation}
from which
\begin{equation}
\phi(t,w)=- \int^{w} \frac{dw^\prime}{\bar N^2(w^\prime,t)}.
\end{equation}
Thus the most general solution of eq.(\ref{eqN3}) has the form
\begin{equation}
\label{gen}
N_{gen.}(w,t)= a(t) \bar N(w,t)+b(t)  \bar N(w,t) \int^{w}
\frac{dw^\prime}{\bar N^2(w^\prime,t)}, 
\end{equation}
where $a(t)$ and $b(t)$ are two arbitrary function depending on time,
but not on $w$. 

As we have seen, a particular solution is provided by   
\begin{eqnarray}
&&N=\frac{\partial (-2\tilde \sigma)}{\partial M}=
\frac{1}{2\pi} \log \left(\frac{r}{\Lambda}\right)-\frac{1}{2\pi(1-\mu)}+
\frac{1}{2\pi}\frac{2}{\left (\frac{r}{\Lambda}\right)^{2(1-\mu)}-1}\log\left 
(\frac{r}{\Lambda}\right)=\nonumber\\
&&=\frac{w}{2\pi}-\frac{1}{2\pi(1-\mu)}+\frac{1}{2\pi}\frac{2w}{ 
e^{2(1-\mu) w}-1},
\end{eqnarray}
which substituted in eq.(\ref{gen}) gives
\begin{equation} 
N_{gen.}(t,w)=a(t)\left [ \frac{w}{2\pi}-\frac{1}{2\pi (1-\mu)}
\frac{b(t)/a(t)}{2\pi (1-\mu)}+ \frac{2}{e^{2(1-\mu)w}-1}\left(w
-\frac{b(t)/a(t)}{2\pi (1-\mu)}\right)\right]. 
\end{equation}
Defining $k(t)\equiv a(t)/b(t)(1-\mu)$ and restoring the variable $r$, the 
above general solution takes the known form
\begin{equation}
N_{gen.}(r,t)=a(t)
\left[\frac{1}{2\pi} \log
\left(\frac{r}{\Lambda\kappa(t)}\right)-\frac{1}{2\pi(1-\mu)}+ 
\frac{1}{2\pi}\frac{2}{\left (\frac{r}{\Lambda}\right)^{2(1-\mu)}-1}\log\left 
(\frac{r}{\Lambda\kappa(t)}\right)\right].
\end{equation}
Requiring that $\displaystyle{N\to \frac{1}{2\pi}\log\left
(\frac{r}{\lambda}\right )}$ when $r$ approaches infinity fixes $a(t)$
to be $1$ and thus we are left with one arbitrary function given by
$k(t)$.

\end{document}